\documentclass[aps,pre,twocolumn]{revtex4}

\newif\ifpdf
\ifx\pdfoutput\undefined
\pdffalse 
\else
\pdfoutput=1 
\pdftrue
\fi

\ifpdf
\usepackage[pdftex]{graphicx}
\else
\usepackage{graphicx}
\fi

\ifpdf
\DeclareGraphicsExtensions{.pdf, .jpg, .tif}
\else
\DeclareGraphicsExtensions{.pdf, .jpg}
\fi

\usepackage{amsfonts}
\usepackage{amsmath}
\usepackage{amssymb}
\usepackage{epsfig}

\def\myhsize{\hsize}
\def\myHsize{2.5in}

\begin{document}
\title{Uniform semiclassical wave function for coherent 2D electron flow}
\author{Ji\v{r}\'{\i} Van\'{\i}\v{c}ek$^1$ and Eric J. Heller$^{1,2}$}
\affiliation{$^1$Department of Physics\\ $^2$Department of Chemistry and Chemical Biology, Harvard University, Cambridge, MA 02138, USA} 

\begin{abstract}

We find a uniform semiclassical (SC) wave function describing coherent branched flow through a two-dimensional electron gas (2DEG), a phenomenon recently discovered by direct imaging of the current
using scanned probed microscopy\cite{TOPINKA,SHAW}.
 The formation of
branches has been explained by classical arguments\cite{SHAW}, but the
SC simulations necessary to account for the coherence are made
difficult by the proliferation of catastrophes in the phase space. In this
paper, expansion in terms of "replacement manifolds" is used to find a uniform
SC wave function for a cusp singularity.  The method is then
generalized and applied to calculate uniform wave functions for a quantum-map
model of coherent flow through a 2DEG.  Finally, the quantum-map
approximation is dropped and the method is shown to work for a continuous-time
model as well.

\end{abstract}

\maketitle

\section{Introduction}

There is no doubt that detailed understanding of the electron transport through mesoscopic devices is needed to take the full advantage of the possibilities of novel electronics these systems offer.  On the experimental side, great progress was made with the use of scanned probe microscopes\cite{ERIKSSON,CROOK,TOPINKA}.  The theory has kept up: the present knowledge has already been summarized in several monographs\cite{SOHN,DATTA,IMRY}.

Quantum effects have become central as devices have become smaller, cooler, and containing fewer impurities. 
Remarkably many quantum properties of the electron flow through nanostructures can be explained by SC methods.  These methods are based on classical mechanics: the relevant classical manifolds form the ``skeleton'' to which the wave function is attached\cite{GUTZWILLER}.
SC methods need to be substituted for classical ones when coherence is maintained over distances on the order of the size of the device, and when interference effects are playing a role.

In their simplest form, the SC techniques fail when nonlinear
classical dynamics create complicated structures in phase space.  In
particular, the SC approximation breaks down whenever there are
multiple contributions to the wave function within the volume of a
single Planck cell.  These so-called catastrophes have been
classified\cite{THOM,ARNOLD} and various methods have been devised to
correct the SC wave functions in cases when there exist only several
coalescing contributions\cite{CARRIER,AIRY_BERRY,PEARCY}. In the
setting of mesoscopic devices, improved SC methods have been applied
e.g. to the 
scattering through ballistic microstructures\cite{BURGDORFER} or
to the magnetotransport through a resonant tunneling diode\cite{STONE}. 

In a recent paper \cite{VANICEK}, we successfully explored a new approach which worked even in situations with an infinite number of coalescing contributions, occurring e.g. in the case of the homoclinic tangle near an unstable periodic orbit\cite{OZORIO}.  This new method is based on the idea of
replacing a complicated classical manifold by a series of new
simpler manifolds. When standard semiclassical methods are applied to these
``replacement manifolds'' (RMs), accurate uniform wave functions are obtained
in situations where direct semiclassical evaluation of the original manifold
fails miserably.

Originally, this method was used in special, although common cases with an
infinite number of oscillations with the same phase-space area. Here we
demonstrate that this special property is not necessary, and that similar
approach may be used more generally, even in cases with localized
perturbations.
In Section~\ref{method}, we briefly review the RM method from Ref. \cite{VANICEK} and generalize it.  The method is used to uniformize a cusp singularity in Section~\ref{cusp_singularity}.  In Section~\ref{electron_flow}, we apply the generalized method to find a uniform wave function
in a quantum-map model of a 2D electron flow through a sample with impurities, where
multiple cusp catastrophes are present.  The quantum-map approximation
is relaxed in Section~\ref{continuous_version} and it is shown how the
replacement manifolds are formed in a continuous-time model.  In
Section~\ref{discussion}, we discuss the merits of the RM method and relate it
to other SC techniques. Because most of this paper is concerned with
what happens to the twisted manifold under the shear of phase space, for
completeness the Appendix addresses the other major phase space motion: rotation.

\section{Replacement-manifold method and its generalization}

\label{method}

The original method, discussed in detail in Ref. \cite{VANICEK}, works for
wave functions of the form
\begin{equation}
\psi(q)=A(q)e^{iS(q)/\hbar}%
\end{equation}
with
\begin{equation}
S(q)=S_{0}(q)+\hbar\epsilon\sin f(q),
\end{equation}
that can be associated with classical manifolds in which the momentum depends
on the position as
\begin{equation}
p(q)=\frac{\partial S}{\partial q}=\frac{\partial S_{0}}{\partial q}%
+\hbar\epsilon f^{\prime}(q)\cos f(q).
\end{equation}
Here $S_{0}$ and $S$ are the unperturbed and full action, respectively,
$\epsilon$ is a parameter controlling the strength of the perturbation, $A(q)
$ gives the local weight of the manifold, and $f(q)$ is a smooth function
defining the shape of the perturbation.

We can expand the wave function as
\begin{equation}
\psi_{RM}(q)=\sum_{n=-\infty}^{\infty}A_{n}(q)\exp\left[  \frac{i}{\hbar}%
S_{n}(q)\right]  , \label{RMexpansion1}%
\end{equation}
and interpret each term of the sum as a contribution from a classical
\textquotedblleft replacement\textquotedblright\ manifold $p_{n}(q)=\partial
S_{n}/\partial q$ with a weight $A_{n}=A(q)J_{n}(\epsilon)$ and an action
$S_{n}(q)=S_{0}(q)+n\hbar f(q)$. The advantage of the RM expansion is
appreciated after moving to the momentum representation with caustics where
semiclassical form $\sum A_{j}^{SC}(p)\exp[iS_{j}^{SC}(p)/\hbar]$ fails while
the sum over RMs gives an accurate result.

A slightly different and more general approach than in Ref. \cite{VANICEK}
does not require an oscillatory behavior of the action. If
\begin{equation}
S(q) = S_{0}(q) + \epsilon\Delta S(q),
\end{equation}
we may Taylor expand the wave function as
\begin{eqnarray}
\psi(q) &=& A(q) \exp\left[  \frac{i}{\hbar} S_{0}(q) \right]  \sum_{n =
0}^{\infty} \frac{1}{n!} \left[  \frac{i}{\hbar} \epsilon\Delta S(q) \right]
^{n} \nonumber \\&=& \sum_{n = 0}^{\infty} A_{n}(q) \exp\left[  \frac{i}{\hbar} S_{n}(q)
\right]  \label{RMexpansion2}%
\end{eqnarray}
corresponding to RMs with weights
\[ A_{n}(q) = A(q) (i \epsilon)^{n} / n!
\] 
and
actions 
\[
S_{n}(q) = S_{0}(q) - i \hbar n \log[\Delta S(q) / \hbar].
\] 
Defining
a new function $f(q)$ by $\Delta S(q) \equiv\hbar\exp[ f(q) ]$, the $n$th RM
action becomes
\begin{equation}
S_{n}(q) = S_{0}(q) - i \hbar n f(q)
\end{equation}
It will ``help'' the convergence of expansion (\ref{RMexpansion2}) if $\lim_{q
\rightarrow\pm\infty} f(q) = -\infty$. This, however, is a natural property of
localized perturbations.

\begin{figure}[htbp]
\centerline{\epsfig{figure=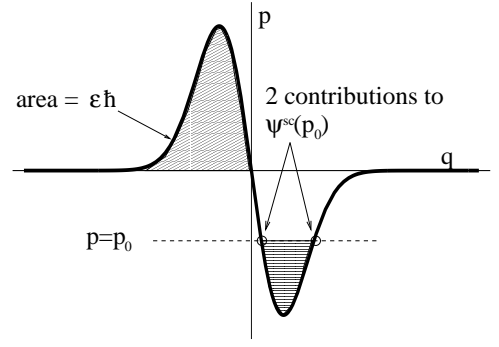,width=\myHsize}}
\caption{Initial manifold and areas important to the semiclassical approximation.} 
\label{loop}
\end{figure}

The simplest nontrivial example is obtained by choosing $f(q)=-q^{2}$. Besides
allowing an analytic solution, this choice will yield exactly the manifold
needed in our model of a 2D electron flow in Section \ref{electron_flow}.
Expanding the function $p(q)$ around $q=0$,
\begin{equation}
p(q)=-2\epsilon\hbar qe^{-q^{2}}\approx2\epsilon\hbar(q^{3}-q)+O(q^{4}),
\label{original_manifold}%
\end{equation}
we find that this case falls into the second simplest universality class
(called cusp) of catastrophe theory \cite{THOM,ARNOLD,BUTTERFLY} (see
Fig.~\ref{loop}).

Assuming that the weighting of this manifold is $A(q)=const.=(2\pi
\hbar)^{-1/2}$, the corresponding SC wave function is
\begin{equation}
\psi_{SC}(q)=(2\pi\hbar)^{-1/2}\,\exp\left(  i\epsilon e^{-q^{2}}\right)  .
\label{psi_large_time}%
\end{equation}
Since for all positions $q$, there exists only a single contribution to
$\psi_{SC}(q)$, the SC position wave function is accurate, $\psi(q)\approx
\psi_{SC}(q)$, and the momentum wave function is given by the Fourier
transform
\begin{equation}
\psi(p)=\left(  2\pi\hbar\right)  ^{-1/2}\int dq\,\psi_{SC}(q)\,e^{-ipq/\hbar
}.
\end{equation}
Evaluating this integral by the stationary-phase (SP) approximation yields the
SC momentum wave function $\psi_{SC}(p)$. The SC momentum wave function has
two contributions from two SP points (Fig.~\ref{loop}). The horizontally
filled-in area gives the phase between two contributions; if it becomes
smaller than $\hbar$, the SP approximation breaks down. Therefore $\psi
_{SC}(p)$ will be singular for all classically allowed momenta when
$\epsilon\leq\hbar$.

Note that the RM momentum $p_{n}(q)=2in\hbar q$ is purely imaginary for all
$q$ and that the corresponding manifold has no caustics. The uniform momentum
wave function is found as
\begin{align}
\psi_{RM}(p)  &  =\left(  2\pi\hbar\right)  ^{-1/2}\int dq\,e^{-ipq/\hbar}%
\sum_{n=0}^{\infty}A_{n}e^{iS_{n}(q)/\hbar}\label{momentum_wf}\\
&  =\delta(p)+\frac{\sqrt{\pi}}{2\pi\hbar}\sum_{n=1}^{\infty}\frac
{(i\epsilon)^{n}}{n!}n^{-1/2}\exp\left(  \frac{-p^{2}}{4n\hbar^{2}}\right)
\nonumber\\
&  =\delta(p)+\sum_{n=1}^{\infty}\tilde{A}_{n}e^{i\tilde{S}_{n}(p)/\hbar}%
\end{align}
where%
\begin{align}
\tilde{A}_{n}  &  =\frac{\sqrt{\pi}}{2\pi\hbar}\frac{(i\epsilon)^{n}}%
{n!}n^{-1/2},\label{momentum_coefficients}\\
\tilde{S}_{n}(p)  &  =-\int dp\,q_{n}(p)=\frac{ip^{2}}{4n\hslash}.
\end{align}
We can in general evaluate all RMs for $n\geq1$ by the SP method, although in this
case the answer turns out to be equal to the exact Fourier transform because
the action $S_{n}$ is quadratic.

\section{Uniformization of a cusp singularity}

\label{cusp_singularity}
The formation of manifolds with a double-loop structure like the one
in Fig.~\ref{loop} is a generic feature of nonlinear Hamiltonian
systems. This pattern forms, for instance, whenever an ensemble of trajectories
encounters a dip or a bump in the potential surface. Assuming that the
particles have energy greater the maximum of the potential, the dip or
bump act as a convex or concave lens, respectively.  After it is
created, the double loop does not remain stationary: depending on the
Hamiltonian, the structure will generally start to shear and rotate in phase
space (see Fig.~\ref{shear_rotation}).  In most of this paper we are concerned with the shear only,
but for completeness, in the Appendix we present analytic formulae for
the RM expansion of an original manifold that is arbitrarily rotated
with respect to the $q$ and $p$ axes. 
\begin{figure}[htbp]

\centerline{\epsfig{figure=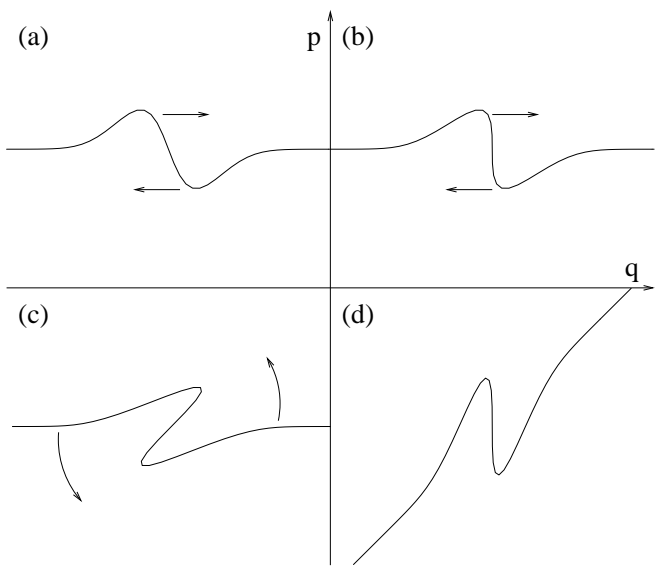,width=\myhsize}}

\caption{An example of a manifold with the double loop structure that has
  been sheared (a-c) and rotated (c-d) in phase space. While shear and rotation are generic phase space motions, here they were implemented by $H=p^2/2$ and $H=p^2/2 + q^2 / 2$, respectively. }
\label{shear_rotation}

\end{figure}
  
\begin{figure}[htbp]

\centerline{\epsfig{figure=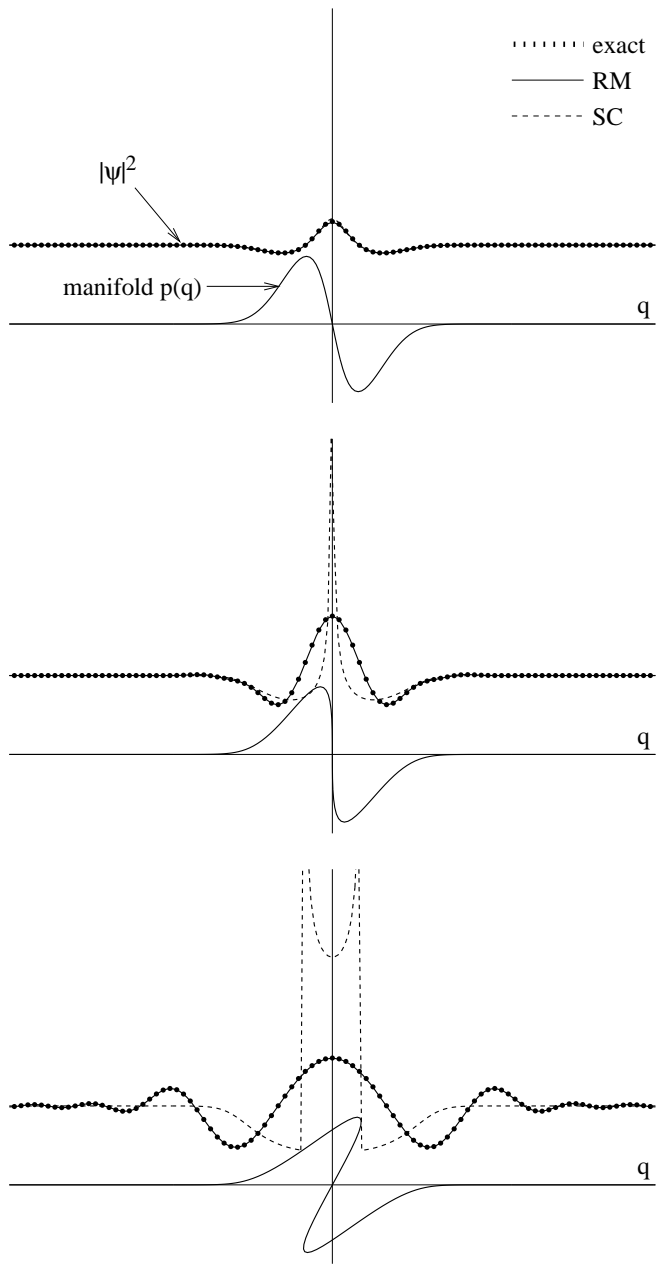,width=\myhsize}}

\caption{Evolution of the manifold and the comparison of the exact
  (points), RM (solid line) and SC (dashed line) wave functions at a
  time instant before ($t=0.25\,t_{\rm cusp}$, top), at ($t=t_{\rm
  cusp}$, middle), and after ($t=3\,t_{\rm cusp}$, bottom) the
  cusp. In these plots, $\epsilon = 1$ and the first five RMs were used.}
\label{wf_for_cusp}

\end{figure}

For now imagine that after the manifold (\ref{original_manifold}) with two loops has been formed, the system
evolves freely (with Hamiltonian $H = p^{2} / 2 m$). Hamilton's equations of
motion are
\begin{align}
\dot{q}  &  = \frac{p}{m},\nonumber\\
\dot{p}  &  = 0,
\end{align}
resulting in a shear of phase-space. The SC position wave function , which was
accurate at time $t = 0$, will break down around time
\begin{equation}
t_{\rm cusp} = \frac{m}{2 \epsilon\hbar}%
\end{equation}
when a cusp singularity\cite{PEARCY} develops (see Fig.~\ref{wf_for_cusp}).

This will be remedied if we apply any of the SC evolution methods (i.e. integration using the SP approximation) to the first
few RMs instead of directly to the original manifold,
\begin{eqnarray}
\psi_{RM}(q,t)&=&\int dq^{\prime}\,K_{f}(q,q^{\prime};t)\,\psi_{RM}(q^{\prime
},0) \\&=& \sum_{n=0}^{\infty} \int dq^{\prime}\,K_{f}(q,q^{\prime};t)\,A_{n}e^{iS_{n}(q)/\hbar} \nonumber
\end{eqnarray}
where the free-space propagator
\begin{equation}
K_{f}(q^{\prime\prime},q^{\prime};t)=\left(  \frac{m}{2\pi\hbar it}\right)
^{1/2}\exp\left[  \frac{im}{2\hbar t}(q^{\prime\prime}-q^{\prime})^{2}\right]
\end{equation}
and at $t=0$, using expression (\ref{RMexpansion2}),
\begin{equation}
\psi_{RM}(q,0)=\left(  2\pi\hbar\right)  ^{-1/2}\sum_{n=0}^{\infty}%
\frac{(i\epsilon)^{n}}{n!}e^{-nq^{2}}. \label{q_representation}%
\end{equation}
In our case, since the RM terms are Gaussian wave packets, their SC evolution (i.e. SP integration)
can be performed analytically and is exact,
\begin{eqnarray}
&&\psi_{RM}(q,t)=\left(  2\pi\hbar\right)^{-1/2} \\ &\times& \sum_{n=0}^{\infty}%
\frac{(i\epsilon)^{n}}{n!}(1+2in\hbar t/m)^{1/2}\exp\left(  \frac{-nq^{2}%
}{1+2in\hbar t/m}\right). \nonumber
\end{eqnarray}

For comparison, the exact quantum evolution was performed by switching to the momentum
representation, using the fast Fourier transform (FFT) and trivially evolving
the wave function there. To find the primitive SC evolution, we used a method
described by Berry et al.\cite{BERRY}. All three methods are compared in
Fig.~\ref{wf_for_cusp}, showing the classical manifold and corresponding
exact, SC, and RM wave function at a time instant before, at, and after the cusp.

In the next Section, we show that the RM method can treat situations in which more cusps are continuously formed.  However, the advantage of the RMs over the Van Vleck propagation or other standard SC methods can be appreciated already when the rough region of the potential is localized in time and only one or a few cusps are created.  As can be seen from Fig.~\ref{wf_for_cusp}, even if the potential is simply flat after certain time, the region of $q$ in which the SC approximation breaks down expands.  Unlike the simple SC approximation which deteriorates with time, the accuracy of the RM method is preserved after leaving the rough area of the potential: once the Gaussian wavepackets corresponding to the RMs are formed, their number remains constant and their propagation is exact in any potential with up to quadratic terms (see Fig.~\ref{wf_for_cusp}).

\section{Quantum-map model of a 2D electron flow through a sample with
impurities}

\label{electron_flow}

\begin{figure}[htbp]

\centerline{\epsfig{figure=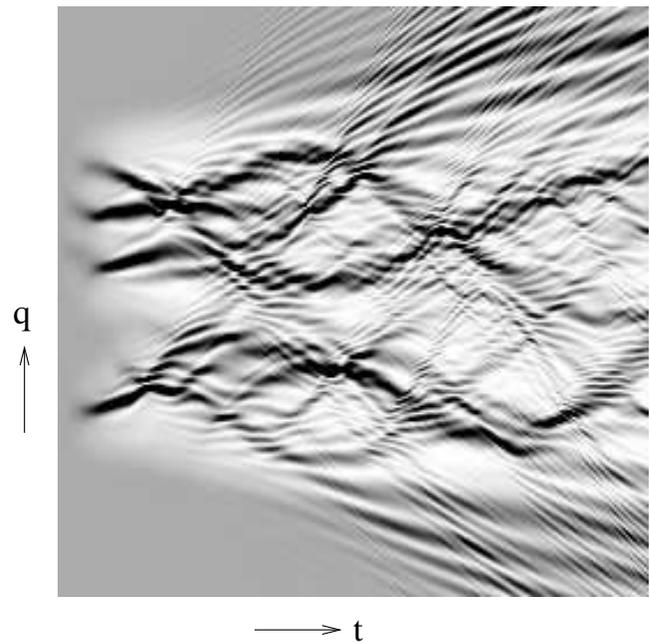,width=\myhsize}}

\caption{Electron density $|\psi(q,t)|^2$ in the model of a 2D electron flow (obtained
  by the exact quantum evolution using the FFT, RMs were not used). In
  this plot, $\epsilon = 2.22$ and there were 256 impurities.}
\label{electron_density}

\end{figure}

\begin{figure}[htbp]

\centerline{\epsfig{figure=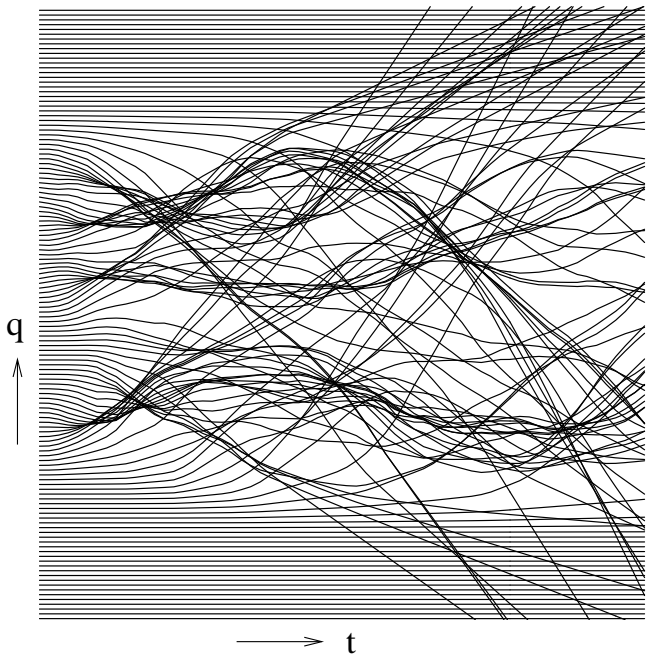 ,width=\myhsize}}

\caption{Representative classical electron trajectories, corresponding
  to the electron density in Fig.~\ref{electron_density}.}
\label{trajectories}

\end{figure}

We are now prepared to address the problem of the 2D electron flow in a semiconductor nanostructure with impurities.  The electron transport in such a system 
is neither strictly ballistic nor strictly diffusive. Instead, the
experiment has revealed that reality lies somewhere in between and the
phenomenon has been termed ``branched flow \cite{SHAW}.''
Figures~\ref{electron_density} and~\ref{trajectories} show
respectively the exact electron density (obtained by the exact quantum
evolution using the FFT) and the representative classical trajectories in the model described below.  The name of the phenomenon comes from the shape of the regions with enhanced electron density in Fig.~\ref{electron_density} or the corresponding clusters of classical electron trajectories in Fig.~\ref{trajectories}. It turns out, however, that these do {\it not}  correspond to the valleys in the potential \cite{SHAW}.  
Although the branches can already be seen in the classical simulations, Fig.~\ref{trajectories} also shows that scattering by impurities leads to
abundant cusp singularities in phase space, and therefore we expect deviations in both
classical and primitive SC approximations from the exact quantum dynamics.

Here we analyze a simple model which can nevertheless exhibit all these
properties. Namely, we discuss a 2D system with fast electrons, incident along
the $x$-axis and scattered by small isolated Gaussian impurities randomly
distributed in the $xy$ plane. Following Topinka et al. \cite{TOPINKA,SHAW}
who observed branched flow in a similar system, we consider the electron
kinetic energy to be much larger than the amplitude of impurities. In fact, we
assume the kinetic energy to be high enough to justify an impulse
approximation: the electron propagates freely between effectively
instantaneous kicks from impurities that affect its momentum but not position.
Moreover, while the transverse momentum changes by a small impulse from an
impurity, the longitudinal momentum remains effectively constant, allowing the
transformation of the original 2D problem into a 1D problem with a
time-dependent Hamiltonian. To be precise, we start with a 2D Hamiltonian
\begin{eqnarray}
&H&(x,y,p_{x},p_{y})=\frac{p_{x}^{2}+p_{y}^{2}}{2m} \\ &+&\sum_{j=1}^{n}V_{0}%
\,\exp\left[  -\frac{(x-x^{(j)})^{2}+(y-y^{(j)})^{2}}{a^{2}}\right] \nonumber
\end{eqnarray}
where $n$, $a$, and $x^{(j)}$, $y^{(j)}$ are respectively the number, radius,
and coordinates of the centers of the impurities. We assume that
\begin{equation}
\dot{x}=p_{x}/m\approx\mathrm{const.}=v
\end{equation}
where $v$ is the initial velocity of the electron. This is justified when
\begin{equation}
mv^{2}/|V_{0}|\gg\sqrt{n}.
\end{equation}
For simplicity of calculations, we distribute the impurities randomly in the
$y$ direction, but regularly along the $x$ axis, at intervals $v\tau$. Each
electron will be affected by a single impurity at a time if
\begin{equation}
v\tau\gg a.
\end{equation}
To simplify notation, we take $a$, $\tau$, and $m$ to be respectively the
units of length, time, and mass. Defining $q=y/a$, our effective, 1D
time-dependent Hamiltonian becomes
\begin{equation}
\label{time_dep_hamiltonian}
H(q,p,t)=\frac{p^{2}}{2}+\sum_{j=1}^{n}V_{0}\exp\left[  -(q-q^{(j)})^{2}%
-v^{2}(t-j)^{2}\right]
\end{equation}
In this section we consider that the change of the transverse momentum due to
the impurity is instantaneous, yielding a further simplification, represented
by a periodically "kicked" Hamiltonian,%

\[
H(q,p,t)\approx\frac{p^{2}}{2}+\sqrt{\pi}\frac{V_{0}}{v}\sum_{j=1}%
^{n}e^{-(q-q^{(j)})^{2}}\delta(t-j).
\]
(A generalized analysis without this approximation is presented in the next
section.) \ In the impulse approximation, classical position $q$ of an
electron does not change during an interaction with $j$th
impurity\cite{IMPULSE}. Classical dynamics may therefore be expressed in terms
of a map,
\begin{align}
q_{j+1}  &  =q_{j}+p_{j},\nonumber\\
p_{j+1}  &  =p_{j}+\Delta p(q_{j+1},q^{(j+1)})
\end{align}
where subscripts denote time in units $\tau$ and the change of momentum is
\begin{equation}
\Delta p(q,q^{(j)})\approx-\int_{-\infty}^{\infty} \!\!dt \,\frac{\partial H}{\partial
q}=2\sqrt{\pi}\frac{V_{0}}{v}(q-q^{(j)})e^{-(q-q^{(j)})^{2}},
\label{momentum_change}%
\end{equation}
implying that a single impurity transforms a momentum state exactly into the
two-loop manifold (\ref{original_manifold}) from Section~\ref{method}. We can
read off the loop area from (\ref{momentum_change}) to be $\sqrt{\pi}%
|V_{0}|/v$.

In quantum mechanics, another important parameter enters: $\hbar$. Accuracy of
the SC approximation will depend on how
\begin{equation}
\epsilon=\sqrt{\pi}\frac{|V_{0}|}{v\hbar}%
\end{equation}
compares to 1. In the impulse approximation, the exact quantum dynamics is
described by a quantum map
\begin{equation}
|\psi_{j+1}\rangle=U\,|\psi_{j}\rangle
\end{equation}
where the subscript again denotes time in units $\tau$ and $U$ is the one-step
evolution operator
\begin{align}
U  &  =T\exp\left(  -\frac{i}{\hbar}\int_{0}^{1}H\,dt\right) \nonumber  \\
& \approx
\exp\left(  -\frac{i}{\hbar}\int_{-\infty}^{\infty}V\,dt\right)  \,\exp\left(
-\frac{i}{\hbar}\frac{p^{2}}{2}\right) \nonumber \\
&  =\exp\left(  i\epsilon\,e^{-q^{2}}\right)  \exp\left(  -\frac{i}{\hbar
}\frac{p^{2}}{2}\right). \label{evolution_op}
\end{align}
The easiest way to evolve a quantum state numerically is to use the FFT to
switch back and forth between position and momentum representations and apply
the impulsive part of $U$ in $q$-representation and the kinetic part of $U$ in
$p$-representation.

We now demonstrate that not only do the replacement manifolds lack
singularities (present in the classical and SC analysis), but that they can
also correctly reproduce all the details of the exact quantum solution. When
the next impurity is encountered, each wave packet develops a loop in its
phase-space representation which would soon lead to a new cusp singularity. We
therefore replace it with a series of simpler manifolds, as in
Section~\ref{cusp_singularity}, avoiding this problem.

In our model we exploit the fact that the RM terms are Gaussian wave packets,
allowing their analytic evaluation with only a slight generalization of the
calculations in Section~\ref{cusp_singularity}. Each term in the RM sum at
time $j$ has a Gaussian form,
\begin{equation}
\psi_{j}(q) = c \, e^{a q - b q^{2}}, \;\;\; \mathrm{Re} \, b > 0.
\label{wavepacket}%
\end{equation}
After the kinetic propagation, just before next impurity is encountered,
\begin{eqnarray}
\tilde{\psi}_{j + 1}(q) &=& \int K_{f}(q, q^{\prime}; 1) \, \psi_{j}(q^{\prime})
\, dq^{\prime} \\ 
&=& \frac{c}{(1 + 2 i b)^{1/2}} \exp\left(  \frac{i a^{2} / 2 + a
q - b q^{2}}{1 + 2 i b} \right). \nonumber
\end{eqnarray}
After receiving an impulse from the $(j + 1)$-st impurity,
\begin{eqnarray}
\psi_{j + 1}(q) &=& \tilde{\psi}_{j + 1}(q) \exp\left[  i \epsilon\, e^{-(q -
q^{(j + 1)})^{2}} \right]  \\
&=& \tilde{\psi}_{j + 1}(q) \sum_{n = 0}^{\infty}
\frac{(i \epsilon)^{n}}{n!} \exp\left[  -n \, (q - q^{(j+1)})^{2} \right]. \nonumber
\end{eqnarray}
Each term in this sum gives rise to a new Gaussian wave packet of the form
(\ref{wavepacket}), which is propagated further in the same manner.

\begin{figure}[htbp]

\centerline{\epsfig{figure=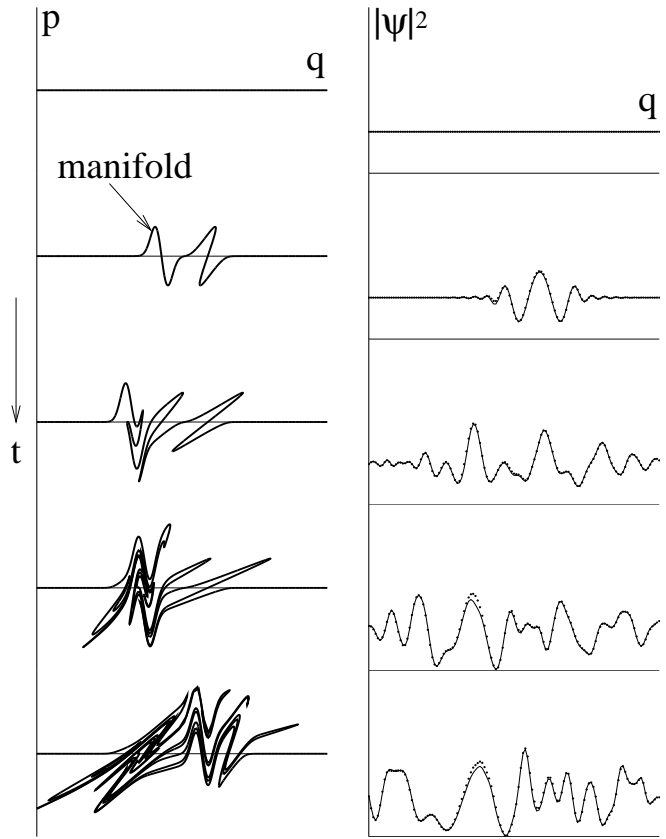,width=\myhsize}}

\caption{Evolution of the manifold (left) and comparison of the
  corresponding exact (points) and RM (solid line) wave functions
  (right). In this plot, $\epsilon = 1.11$ and the first four RMs were
  used.  Two new impurities were encountered at each time interval
  between the consecutive rows.} 
\label{wf_for_impurities}

\end{figure}

Figure~\ref{wf_for_impurities} shows a comparison of the exact and RM
evolution for $\epsilon = 1.11$ up to a time when eight impurities are
encountered. Four RMs are used to replace each incident wave packet at each
impurity. Although the classical manifold (also shown in the figure) has
developed many structures smaller than $\hbar$, the agreement remains excellent.

\section{Continuous version of the model}

\label{continuous_version}

In certain situations, we may be interested in a detailed evolution of the
electron wave function during the collision with the impurity, rather than
just in the appearance of the wave function after the collision. Below, we
present an analytical solution of this problem in case\ that the electrons
move slowly enough that the collision cannot be considered instantaneous, but
fast enough that the transverse displacement of the electrons does not change
significantly during the collision. (For even slower electrons, the
coupling between the longitudinal and transverse motion during the
collision would prevent us from obtaining closed analytic expressions
presented below.  However, we could still find the replacement
manifolds numerically.) 

To simplify the notation, we consider only
a single impurity located at position $q=0$ and time $t=0$, so that the
effective 1D time-dependent Hamiltonian (\ref{time_dep_hamiltonian}) becomes%
\begin{equation}
H(q,p,t)=\frac{1}{2}p^{2}+V_{0}\exp\left(  -q^{2}-v^{2}t^{2}\right)  .
\end{equation}
Assuming that $q$ changes little during the collision we find that
the momentum change is%
\begin{eqnarray}
\Delta p &=& p(t)-p(-\infty)=\int_{-\infty}^{t}dt^{\prime}\,\dot{p}(t^{\prime
})=-\int_{-\infty}^{t}dt^{\prime}\,\frac{\partial H}{\partial q} \nonumber \\
&=&2qV_{0}e^{-q^{2}}\int_{-\infty}^{t}dt^{\prime}\,\exp\left(  -v^{2}t^{\prime
2}\right)  \nonumber \\
&=& \sqrt{\pi}V_{0}v^{-1}qe^{-q^{2}}\left[  1+\operatorname{erf}%
\left(  vt\right)  \right]
\end{eqnarray}
At time $t=-\infty$, we start with a momentum eigenstate with momentum $p=0$,%
\begin{align}
\psi(p,t  &  =-\infty)=\delta(p),\label{psi_small_time}\\
\text{or }\psi(q,t  &  =-\infty)=\left(  2\pi\hbar\right)  ^{-1/2}\nonumber
\end{align}
represented by a horizontal line in phase space. As the electron wave passes
through the impurity, a double loop develops in the manifold (curve)
representing the wave function. The position representation of the SC wave
function at time $t$ is (see, e.g. \cite{BERRY})%
\[
\psi(q_{f},t)=\left(  2\pi\hbar\right)  ^{-1/2}\left\vert \frac{dq_{i}}%
{dq_{f}}\right\vert ^{1/2}\exp\left[  \frac{i}{^{\hbar}}\left(  S_{1}%
+S_{2}\right)  \right]  ,
\]
where $q_{i}$ is the position at time $t^{\prime}=-\infty$ that evolves to
position $q_{f}$ at time $t^{\prime}=t$. In our approximation $q_{f}$ $\approx
q_{i}$, the Van Vleck determinant $|dq_{i}/dq_{f}|=1$, which is the reason
that the SC position wave function remains accurate throughout the collision.
$S_{1}$ is the action along the trajectory of a reference point $q=x$ on the
manifold,%
\[
S_{1}=\int_{-\infty}^{t}dt^{\prime}L\left[  x(t^{\prime}),\dot{x}(t^{\prime
}),t^{\prime}\right]  .
\]
$S_{2}$ is the reduced action along the evolved manifold at time $t$,%
\[
S_{2}=\int_{x_{f}}^{q_{t}}dq_{t}^{\prime}\,p_{t}\left(  q_{t}^{\prime}\right)
\]
($p_{t}(q)$ is the momentum dependence on position at time $t$). For
convenience, we choose $x(-\infty)=-\infty$, giving $\dot{x}(t)=0$ and
$x(t)=const.=-\infty$. Since $V(x=-\infty,t)=0$, also $L(x,\dot{x},t)=0$ and
$S_{1}=0$. Finally, since $p_{t=-\infty}(q)=0$,%
\begin{eqnarray}
&&S_{2}=\int_{-\infty}^{q}dq^{\prime}\,\Delta p_{t}(q^{\prime})=\sqrt{\pi}%
V_{0}v^{-1}\left[  1+\operatorname{erf}\left(  vt\right)  \right] \nonumber \\
&\times&\int_{-\infty}^{q}\!\!dq^{\prime}\,q^{\prime}\,e^{-q^{\prime2}}=\sqrt{\pi}%
V_{0}v^{-1}\frac{1}{2}\left[  1+\operatorname{erf}\left(  vt\right)  \right]
\,e^{-q^{2}}%
\end{eqnarray}
The semiclassical position wave function at time $t$ is%
\begin{eqnarray}
\psi_{SC}(q,t)&=&\left(  2\pi\hbar\right)^{-1/2} \nonumber \\ \times \exp\left\{  \frac{i}{\hbar
}\sqrt{\pi}V_{0}v^{-1}\frac{1}{2}\left[  1+\operatorname{erf}\left(
vt\right)  \right]  \,e^{-q^{2}}\right\}  \label{psi_continuous_time}%
\end{eqnarray}
Remembering that $\sqrt{\pi}V_{0}v^{-1}\hbar^{-1}=\epsilon$ and that
$\operatorname{erf}\left(  \pm\infty\right)  =\pm\infty$, we can easily check
that this general expression gives the correct limiting forms
(\ref{psi_small_time}) and (\ref{psi_large_time}) at times $t=-\infty$ and
$t=\infty$, respectively. \ The primitive SC momentum wave function (obtained
by the SPA of the Fourier transform of (\ref{psi_continuous_time})) fails for
the same reasons as in Section \ref{method}. If we expand $\psi_{SC}(q,t)$ in
terms of RMs, and apply the SPA directly to the RMs, we find an accurate
answer. The only difference from expression (\ref{momentum_wf}) is an extra
factor $\left\{  \frac{1}{2}\left[  1+\operatorname{erf}\left(  vt\right)
\right]  \,\right\}  ^{n}$ for RM coefficients $A_{n}$ or $\tilde{A}_{n}$
(\ref{momentum_coefficients}), e.g.%

\[
\tilde{A}_{n}=\frac{\sqrt{\pi}}{2\pi\hbar}\frac{(i\epsilon)^{n}}{n!}%
n^{-1/2}\left\{  \frac{1}{2}\left[  1+\operatorname{erf}\left(  vt\right)
\right]  \,\right\}  ^{n}.
\]
Since the expression in the large parentheses goes smoothly from 0 at
$t=-\infty$ to 1 at $t=\infty$, we see that the replacement manifolds emerge
even before the center of the impurity is encountered. \ However, the weight
of the manifolds with larger $n$ becomes appreciable only after the impurity
is passed.

\section{Discussion and Conclusions}

\label{discussion}

We have shown that the RM method is not limited to infinitely repeating
phase-space structures if we allow the replacement manifolds to have complex
momenta. Propagation of replacement manifolds gives uniform semiclassical wave
functions long after the primitive semiclassical approximation breaks down.

Putting aside the accuracy, the RM approach may seem intimidating from a
numerical point of view because as described, the algorithm has exponential
complexity. But let us remember that the same--exponential proliferation of
contributions--is true of the primitive semiclassical solution--which,
however, would give a completely wrong result in our case! Moreover, there
appear to be at least two possible ways to speed up the RM calculations: For
$\epsilon<1$ , we could prune the contributions to keep only terms up to a
certain \textquotedblleft total\textquotedblright\ power of $\epsilon$ (which
is different from keeping all terms up to a given power at each impurity). Or
we could consolidate the number of wavepackets after certain time by projecting on a
suitable basis (because the exponentially growing number of RM terms is
obviously over-complete) and starting the RM propagation afresh. 

The
question of computational complexity would not even arise if we were interested in a
system where the electron wave hits only one or a few impurities and
after that propagates in a relatively smooth potential.  The small number of Gaussian
wavepackets spawned at the last impurity would suffice for all
subsequent times and the accuracy of the approximation would be preserved. 
As discussed in Section~\ref{cusp_singularity}, this should be contrasted with the standard SC approximation which deteriorates even when a manifold with a single cusp propagates in a flat potential (see Fig.~\ref{wf_for_cusp}). 

Besides providing a uniform wave function the RM method gives an
intuitive explanation of how quantum mechanics smooths out the
classical detail.  Moreover, in the present case of RMs with a complex
momentum, the method appears to provide a link between the
semiclassical perturbation approximations\cite{MILLER} and various
Gaussian wavepacket techniques\cite{FROZEN,HERMAN_KLUK,SPAWNING},
because replacement manifolds in the expansion
(\ref{q_representation}) are nothing but Gaussian wavepackets.  One
advantage of the RM method over other Gaussian wavepacket methods lies
in that it gives an analytic expression for the coefficients of the
wavepackets.  Other Gaussian wavepacket methods (such as the frozen
Gaussians \cite{FROZEN}, the Herman-Kluk
propagator \cite{HERMAN_KLUK} and the full multiple spawning
\cite{SPAWNING}) rely on variational or ad hoc methods to obtain optimal
wavepacket coefficients numerically. 

Finally, although the RM method has not yet been fully generalized, the large variety of problems (in this paper and in Ref.~\cite{VANICEK}) it can solve suggests that the method (or at least, {\it the idea}) is more general.

\section{Acknowledgments}

This research was supported by the National Science Foundation under Grant No.
CHE-0073544 and by the Institute for Theoretical Atomic and Molecular Physics. One of
us (J. V.) would like to acknowledge helpful discussions with D. Cohen and A. Mody.

\appendix

\section{Replacement-manifold expansion in an arbitrarily rotated coordinate
system}

\begin{figure}[htbp]

\centerline{\epsfig{figure=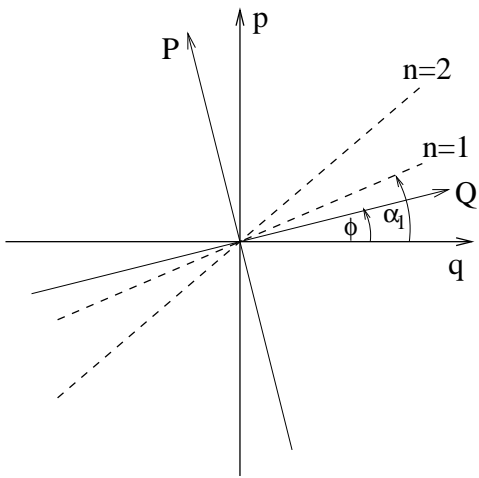,width=\myHsize}}

\caption{Original and rotated coordinate systems.}
\label{rotated_frame}

\end{figure}

We show here that the RM expansion for the manifold studied in this paper can
be found analytically in an arbitrarily rotated coordinate system. In other
words, the method can be readily applied not just in the $q$ or $p$
representations, but in any mixed representation given by a canonical
transformation (see Fig.~\ref{rotated_frame})
\begin{align}
Q  &  =q\cos\phi+p\sin\phi\nonumber\\
P  &  =-q\sin\phi+p\cos\phi
\end{align}

In the original coordinate system ($q$, $p$), replacement manifolds are
straight lines
\begin{equation}
p_{n}(q)=2in\hbar q=q\tan\alpha_{n}%
\end{equation}
(where $\alpha_{n}$ is complex). In the rotated coordinates ($Q$, $P$), the
replacement manifolds are given by the relationship
\begin{equation}
P_{n}(Q)=Q\tan(\alpha_{n}-\phi)=Q\frac{2in\hbar-\tan\phi}{1+2in\hbar\tan\phi}.
\end{equation}
The reduced action is
\begin{equation}
S_{n}(Q)=\int dQ\,P_{n}(Q)=\frac{1}{2}Q^{2}\tan(\alpha_{n}-\phi).
\end{equation}
The weight of the $n$th RM in the $Q$ representation is the weight in the $q$
representation multiplied by the ratios of the projections on the $q$ and $Q$
axes, respectively. Including the Maslov index $\mu$ (0 or 1), we find the
correct $n$th RM contribution
\begin{eqnarray}
\psi_{RM,n}(Q)&=&\frac{1}{\sqrt{2\pi\hbar}}\frac{(i\epsilon)^{n}}{n!}\left\vert
\frac{\cos\alpha_{n}}{\cos(\alpha_{n}-\phi)}\right\vert ^{1/2} \nonumber \\ 
&\times& \exp\left[\frac{i}{2\hbar}Q^{2}\tan(\alpha_{n}-\phi)-i\mu_{n}\pi/2\right]  .
\end{eqnarray}
After simplification, the full RM expansion becomes
\begin{align}
\psi_{RM}(Q)  &  =\sum_{n=0}^{\infty}\psi_{RM,n}(Q)\\
&  =\frac{1}{\sqrt{2\pi\hbar}}\sum_{n=0}^{\infty}\frac{(i\epsilon)^{n}}%
{n!}|\cos\phi+2in\hbar\sin\phi|^{-1/2} \nonumber \\
& \times \exp\left(  -Q^{2}\frac{n+i\tan
\phi/(2\hbar)}{1+2in\hbar\tan\phi}-i\mu_{n}\pi/2\right)  .\nonumber
\end{align}
It is easy to check that this general result correctly reduces to expressions
(\ref{momentum_wf}) or (\ref{q_representation}) when $\phi=\pi/2$ or
$\phi=0$, respectively.


\begin{thebibliography}{99}
\bibitem{BUTTERFLY} Another choice, $f(q) = -\frac{1}{4} q ^4 + \frac{1}{2} q^2$, would
give an example of another universality class: the so-called butterfly
catastrophe.
\bibitem{IMPULSE}  With this approximation, the only
classical parameter remaining is $V_0 / v$.
\end{thebibliography}
\end{document}